\documentstyle[12pt,epsfig]{article}
\def\bge{\begin{equation}}
\def\ene{\end{equation}}
\def\bg{\begin{eqnarray}}
\def\en{\end{eqnarray}}
\def\non{\nonumber}

\def\ubar{\bar{u}}

\def\cbar{\bar{c}}
\def\qbar{\bar{q}}
\def\Qbar{\overline{Q}}
\def\Kbar{\overline{K}}
\def\Dbar{\overline{D}}
\def\D0bar{\overline{D^0}}
\def\Bbar{\overline{B}}

\begin{document}
\begin{titlepage}
\title{Properties of charmed and bottom hadrons\\ in nuclear matter:
A plausible study}
\author{
K. Tsushima$^1$~\thanks{tsushima@physast.uga.edu}~,
F.C. Khanna$^2$~\thanks{Permanent address: 
Department of Physics, University of Alberta, Edmonton, Canada, T6G 2J1
\newline\hspace*{1.5em} khanna@Phys.UAlberta.CA}
\\ \\
{$^1$\small Department of Physics and Astronomy,
University of Georgia, Athens, GA 30602, USA} \\
{$^2$\small CSSM, University of Adelaide, Adelaide, SA 5005, Australia}
}
\maketitle
\vspace{-9cm}
\hfill Alberta Thy 11-02
\vspace{9cm}
\begin{abstract}
Changes in properties of heavy hadrons with a charm or 
a bottom quark are studied in nuclear matter.
Effective masses (scalar potentials) for the hadrons are calculated using 
quark-meson coupling model. 
Our results also suggest that the heavy baryons containing a charm 
or a bottom quark will form charmed or bottom hypernuclei, 
which was first predicted in mid 70's.
In addition a possibility of $B^-$-nuclear bound 
(atomic) states is briefly discussed.
\\ \\
{\it PACS number(s)}: 24.85, 14.20.L, 14.20.M, 21.65, 11.30.R  
\end{abstract}
\end{titlepage}
%
   Extensive studies with hypernuclei have been carried out over 
the last 20 years~\cite{hyplit,hypexp}. 
These involve embedding a
$\Lambda$-particle (hyperon), 
with one (or two) strange quark (quarks) combined 
with u and (or) d quarks (quark),  
in finite nuclei and then studying the single 
particle states, spin-orbit interaction and finally the overall 
binding of the particle in nuclei with different A, number of 
ordinary baryons, nucleons, n and p. Such studies have been hindered since 
there has been no high intensity source of kaon
beams that interact with nuclei to produce $\Lambda$-particles. 

Recently theoretical studies have been extended to 
take account of the quark structure of the 
baryons~\cite{Tsushima_hyp,Oka,Nemura}. 
Agreement with sparse experimental data~\cite{hypexp} 
is impressive. Lately 
there have been attempts to look for a bound state of 
6-quarks, the so-called H particle predicted by Jaffe~\cite{Jaffe}, 
with no success~\cite{Hparticle}. There has been confirmation of 
a bound state of two $\Lambda$-particles   
to a finite nucleus (double hypernucleus)~\cite{doublehyp}. 
All these experimental and theoretical studies were 
directed to learn about the hadrons containing strange quarks 
in a surroundings of nuclear sea made of mainly valence u and d quarks, 
although probably there are no quark studies for 
the double hypernucleus up to now, in spite of its importance and 
recent experimental achievements. 

The approved construction of the Japan Hadron Facility (JHF) will 
be essentially a kaon factory, thus it is expected to produce large 
fluxes of hyperons that should allow a detailed study of hypernuclei. 
However, the facility will be much more than a 
kaon factory. With a beam energy of 50 GeV, it will produce charmed hadrons 
profusely and bottom hadrons in lesser 
numbers but still with an intensity that is comparable  
to the present hyperon production rates.
In mid 70's, a possible formation of the charmed hypernuclei 
were predicted theoretically~\cite{Tyapkin,Dover}. There was an 
experimental search of the charmed and bottom hypernuclei 
at the ARES facility~\cite{charmexp1},  
and it was also investigated 
at the possible $c\tau$-factory~\cite{charmexp2}. 
It is clear that situation for the experiments to search for such 
charmed and bottom hypernuclei is now becoming realistic and would be 
realized at JHF.

This brings us to initiate a careful study of nuclei with 
charm or bottom quarks. 
The production of 
charmonium ($\cbar c$), mesons with charm, 
and baryons with charm quarks will be 
sufficiently large to make it possible to 
study charmed hypernuclei. Study of such nuclei would initially 
involve single particle energies, spin-orbit interaction 
and overall binding energies. Studies with a charm quark and 
a bottom quark in a many-body system would provide 
the first opportunity to learn about the behavior of hadrons 
containing heavy quarks in a sea of valence u and d quarks.
Eventually a study of 
the decay of such hadrons will be a valuable lesson in finding 
the effect of many-body systems on the intrinsic 
properties of charmed and bottom hyperons.
The advantage of using hadrons with heavy quarks  
is that they can convey an information at 
short distance, i.e., that of the very central region of the nucleus 
from charmed and bottom hypernuclei. 
Meson nuclear atomic bound states provide useful information 
about the surface of the nucleus. 

The present investigation is devoted to a study of baryons (and mesons)
which contain a charm or a bottom quark (will be denoted by $C$) 
in nuclear matter. 
Although the baryons with a charm or a bottom quark which we wish to 
study have a typical mean life of the order $10^{-12}$ seconds  
(magnitude is shorter than hyperons), we would like to
gain an understanding of the movement of such a hadron in its 
nucleonic environment.
This would lead to an effective mass (scalar potential) for 
the hadron. The light quark in the hadron (and nucleons) would 
change its property in nuclear medium in a self-consistent manner, and 
will thus affect the overall interaction with nucleons.
With this understanding we will be in a better 
position to learn about the hadron properties with the presence of heavy 
quarks, or baryons with heavy quarks in finite nuclei that will be 
the real ground for these experimental studies.

      At JHF, in addition to charmed  and bottom hyperons, mesons with open
charm (bottom) like $D^- (\bar{c}d)$ ($B^- (\bar{u}b)$) will be 
produced. Such mesons like $K^- (\bar{u}s)$ can form mesic atoms around finite
nuclei. The atomic orbits will be very 
small and will thus probe the surface of light nuclei and will be within the
charge radii for heavier nuclei. Thus at 
least for light nuclei they will give a precise information about the charge
density.

Furthermore, in considering recent experimental situation on 
high energy heavy ion collisions, to study general properties of heavy 
hadrons in nuclear medium is useful, because  
elementary hadronic reactions 
occur in high nuclear density zone of the collisions, and 
many hadrons produced there are under effects of a surrounding nuclear medium.
Thus, we need to understand the properties of heavy hadrons  
in nuclear medium. Some such applications were also made for 
$J/\Psi$ dissociation in nuclear matter, and 
$D$ and $\Dbar$ productions in antiproton-nucleus 
collisions~\cite{Alexd}.

At present we need to resort to a model which can 
describe the properties of finite nuclei as well as hadron properties 
in nuclear medium based on the quark degrees of freedom. 
Although some studies for heavy mesons with charm 
in nuclear matter were made by QCD sum rule for 
$J/\Psi$~\cite{Hayashigaki1,Klingl} and $D (\Dbar)$~\cite{Hayashigaki2} 
there seems to exist no studies for heavy baryons with a charm  
or a bottom quark.
With its simplicity and applicability, we use 
quark-meson coupling (QMC) model~\cite{Guichon}, which has been 
extended and successfully applied to many  
problems in nuclear 
physics~\cite{Guichonf,Saitof,qmc2,qmcf,Blunden,Jin,qmcnuc} 
including a detailed study of the properties of 
hypernuclei~\cite{Tsushima_hyp}, and harmonic 
properties in nuclear 
medium~\cite{Alexd,Tsushimak,qmcapp,Tsushimad}.
In particular, recent measurements of polarization transfer performed 
at MAMI and Jlab~\cite{JlabQMC} support the medium modification of the 
proton electromagnetic form factors calculated by the QMC model. 
The final analysis~\cite{Strauch} 
seems to become more in favor of QMC, although   
still error bars may be large to draw a definite conclusion.
This gives us confidence that such a quark-meson coupling model will provide
us with valuable glimpse into the 
properties of charmed- and bottom-hypernuclei.

We start to consider static, (approximately) spherically symmetric 
charmed and bottom hypernuclei (closed shell plus one heavy baryon 
configuration) ignoring small nonspherical 
effects due to the embedded heavy baryon. 
We adopt Hartree, mean-field, approximation.  
In this approximation, $\rho NN$ tensor coupling gives 
a spin-orbit force for a nucleon bound 
in a static spherical nucleus, although 
in Hartree-Fock it can give a central force which contributes to 
the bulk symmetry energy~\cite{Guichonf,Saitof}. 
Furthermore, it gives no contribution for nuclear 
matter since the meson fields are independent of position 
and time. Thus, we ignore the $\rho NN$ tensor coupling  
as usually adopted in the Hartree treatment of 
quantum hadrodynamics (QHD)~\cite{QHD1,QHD2}.

Using the Born-Oppenheimer approximation, mean-field equations
of motion are derived for a charmed (bottom) hypernucleus  
in which the quasi-particles moving
in single-particle orbits are three-quark clusters with the quantum numbers
of a charmed (bottom) baryon or a nucleon. 
Then a relativistic
Lagrangian density at the hadronic 
level~\cite{Guichonf,Saitof} can be constructed, 
similar to that obtained in QHD~\cite{QHD1,QHD2},
which produces the same equations of motion
when expanded to the same order in velocity:
\begin{eqnarray}
{\cal L}^{CHY}_{QMC} &=& {\cal L}_{QMC} + {\cal L}^C_{QMC}, 
\non \\
{\cal L}_{QMC} &=&  \overline{\psi}_N(\vec{r}) 
\left[ i \gamma \cdot \partial
- M_N^{\star}(\sigma) - (\, g_\omega \omega(\vec{r}) 
+ g_\rho \frac{\tau^N_3}{2} b(\vec{r}) 
+ \frac{e}{2} (1+\tau^N_3) A(\vec{r}) \,) \gamma_0 
\right] \psi_N(\vec{r}) \quad \non \\
  &-& \frac{1}{2}[ (\nabla \sigma(\vec{r}))^2 +
m_{\sigma}^2 \sigma(\vec{r})^2 ]
+ \frac{1}{2}[ (\nabla \omega(\vec{r}))^2 + m_{\omega}^2
\omega(\vec{r})^2 ] \non \\
 &+& \frac{1}{2}[ (\nabla b(\vec{r}))^2 + m_{\rho}^2 b(\vec{r})^2 ]
+ \frac{1}{2} (\nabla A(\vec{r}))^2, \non \\
{\cal L}^C_{QMC} &=& \sum_{C=\Lambda_c,\Sigma_c,\Xi_c,\Lambda_b}     
\overline{\psi}_C(\vec{r}) 
\left[ i \gamma \cdot \partial
- M_C^{\star}(\sigma)
- (\, g^C_\omega \omega(\vec{r}) 
+ g^C_\rho I^C_3 b(\vec{r}) 
+ e Q_C A(\vec{r}) \,) \gamma_0 
\right] \psi_C(\vec{r}), \qquad \label{Lagrangian}
\end{eqnarray}
where $\psi_N(\vec{r})$ ($\psi_C(\vec{r})$)
and $b(\vec{r})$ are respectively the
nucleon (charmed and bottom baryon) and the $\rho$
meson (the time component in the third direction of
isospin) fields, while $m_\sigma$, $m_\omega$ and $m_{\rho}$ are
the masses of the $\sigma$, $\omega$ and $\rho$ meson fields.
$g_\omega$ and $g_{\rho}$ are the $\omega$-$N$ and $\rho$-$N$
coupling constants which are related to the corresponding
(u,d)-quark-$\omega$, $g_\omega^q$, and
(u,d)-quark-$\rho$, $g_\rho^q$, coupling constants as
$g_\omega = 3 g_\omega^q$ and
$g_\rho = g_\rho^q$~\cite{Guichonf,Saitof}.
(See also Eqs.(\ref{diracu}) and (\ref{diracd}).) 
Note that in usual QMC (QMC-I) 
the meson fields appearing 
in Eq.~(\ref{Lagrangian}) represent the quantum numbers and 
Lorentz structure as those used in QHD~\cite{QHD2},
corresponding, $\sigma \leftrightarrow \phi_0$, 
$\omega \leftrightarrow V_0$ and $b \leftrightarrow b_0$, 
and they are not directly connected with the  
physical particles, nor quark model states. 
Their masses in nuclear medium do not vary 
in the present treatment.
For the other version of QMC (QMC-II), where masses of the meson fields 
are also subject to the medium modification in a self-consistent 
manner, see Ref.~\cite{qmc2}. However, for a proper 
parameter set (set B) the typical results obtained 
in QMC-II are very similar to those of QMC-I.
The difference is $\sim 16$ \% for the largest case, but  
typically $\sim 10$ \% or less. (For the effective 
masses of the hyperons, it is less than $\sim 8$ \%.)

In an approximation where the $\sigma$, $\omega$ and $\rho$ fields couple
only to the $u$ and $d$ quarks,  
the coupling constants in the charmed (bottom) baryon 
are obtained as $g^C_\omega = (n_q/3) g_\omega$, and
$g^C_\rho = g_\rho = g_\rho^q$, with $n_q$ being the total number of
valence $u$ and $d$ (light) quarks in the baryon $C$. $I^C_3$ and $Q_C$
are the third component of the baryon isospin operator and its electric
charge in units of the proton charge, $e$, respectively.
The field dependent $\sigma$-$N$ and $\sigma$-$C$
coupling strengths predicted by the QMC model,
$g_\sigma(\sigma)$ and  $g^C_\sigma(\sigma)$,
related to the Lagrangian density, 
Eq.~(\ref{Lagrangian}), at the hadronic level are defined by:
\bg
M_N^{\star}(\sigma) &\equiv& M_N - g_\sigma(\sigma)
\sigma(\vec{r}) ,  \\
M_C^{\star}(\sigma) &\equiv& M_C - g^C_\sigma(\sigma)
\sigma(\vec{r}) , \label{coupny}
\en
where $M_N$ ($M_C$) is the free nucleon (charmed and bottom baryon) 
mass (masses).
Note that the dependence of these coupling strengths on the applied
scalar field must be calculated self-consistently within the quark
model~\cite{Tsushima_hyp,Guichonf,Saitof}. 
Hence, unlike QHD~\cite{QHD1,QHD2}, even though
$g^C_\sigma(\sigma) / g_\sigma(\sigma)$ may be
2/3 or 1/3 depending on the number of light quarks in the baryon 
in free space ($\sigma = 0$)\footnote{Strictly, this is true
only when the bag radii of nucleon and 
heavy baryon $C$ are exactly the same
in the present model. See Eq.~(\ref{hmass}), below.}, 
this will not necessarily  be the case in
nuclear matter.

In the following, we consider the system in the limit of 
infinitely large, uniform (symmetric) nuclear matter, 
where all scalar and vector fields become constants.
Furthermore, under this 
limit, we may also treat a hadron $h$  
embedded in the nuclear matter system, 
in the same way as that for the charmed (bottom) baryon.
(A Lagrangian density for a meson-nuclear system can be also written 
in a similar 
way to that of the charmed (bottom) hypernuclei system, if 
${\cal L}^C_{QMC}$ is replaced by the corresponding meson Lagrangian 
density in Eq.(\ref{Lagrangian}).)

Then, the Dirac equations for the quarks and antiquarks 
in nuclear matter, in bags of 
hadrons, $h$, ($q = u$ or $d$, and $Q = s,c$ or $b$, hereafter) 
neglecting the Coulomb force in nuclear matter, are given by  
($|\mbox{\boldmath $x$}|\le$ bag 
radius)~\cite{Tsushimak,qmcapp,Tsushimad}:
\begin{eqnarray}
\left[ i \gamma \cdot \partial_x -
(m_q - V^q_\sigma)
\mp \gamma^0
\left( V^q_\omega +
\frac{1}{2} V^q_\rho
\right) \right] 
\left( \begin{array}{c} \psi_u(x)  \\
\psi_{\bar{u}}(x) \\ \end{array} \right) &=& 0,
\label{diracu}\\
\left[ i \gamma \cdot \partial_x -
(m_q - V^q_\sigma)
\mp \gamma^0
\left( V^q_\omega -
\frac{1}{2} V^q_\rho
\right) \right]
\left( \begin{array}{c} \psi_d(x)  \\
\psi_{\bar{d}}(x) \\ \end{array} \right) &=& 0,
\label{diracd}\\
\left[ i \gamma \cdot \partial_x - m_{Q} \right]
\psi_{Q} (x)\,\, ({\rm or}\,\, \psi_{\Qbar}(x)) &=& 0. 
\label{diracQ}
\end{eqnarray}
The (constant) mean-field potentials for a bag in nuclear matter
are defined by $V^q_\sigma \equiv g^q_\sigma \sigma$, 
$V^q_\omega \equiv g^q_\omega \omega$ and
$V^q_\rho \equiv g^q_\rho b$,
with $g^q_\sigma$, $g^q_\omega$ and
$g^q_\rho$ the corresponding quark-meson coupling
constants. 

The normalized, static solution for the ground state quarks or antiquarks
with flavor $f$ in the hadron, $h$, may be written,  
$\psi_f (x) = N_f e^{- i \epsilon_f t / R_h^*}
\psi_f (\mbox{\boldmath $x$})$,
where $N_f$ and $\psi_f(\mbox{\boldmath $x$})$
are the normalization factor and
corresponding spin and spatial part of the wave function. 
The bag radius in medium for a hadron $h$, $R_h^*$, 
will be determined through the
stability condition for the mass of the hadron against the
variation of the bag 
radius~\cite{Guichon,Guichonf,Saitof}
(see Eq.~(\ref{hmass})). 

The eigenenergies  
in units of $1/R_h^*$ are given by, 
\bge
\left( \begin{array}{c}
\epsilon_u \\
\epsilon_{\bar{u}}
\end{array} \right)
= \Omega_q^* \pm R_h^* \left(
V^q_\omega
+ \frac{1}{2} V^q_\rho \right),\,\,
\left( \begin{array}{c} \epsilon_d \\
\epsilon_{\bar{d}}
\end{array} \right)
= \Omega_q^* \pm R_h^* \left(
V^q_\omega
- \frac{1}{2} V^q_\rho \right),\,\,
\epsilon_{Q}
= \epsilon_{\Qbar} =
\Omega_{Q}.
\label{energy}
\ene

The hadron masses
in a nuclear medium $m^*_h$ (free mass will be denoted by $m_h$),
are calculated by
\begin{eqnarray}
m_h^* &=& \sum_{j=q,\bar{q},Q,\Qbar} 
\frac{ n_j\Omega_j^* - z_h}{R_h^*}
+ {4\over 3}\pi R_h^{* 3} B,\quad
\left. \frac{\partial m_h^*}
{\partial R_h}\right|_{R_h = R_h^*} = 0,
\label{hmass}
\end{eqnarray}
where $\Omega_q^*=\Omega_{\bar{q}}^*
=[x_q^2 + (R_h^* m_q^*)^2]^{1/2}\,(q=u,d)$, with
$m_q^*=m_q{-}g^q_\sigma \sigma$,
$\Omega_Q^*=\Omega_{\Qbar}^*=[x_Q^2 + (R_h^* m_Q)^2]^{1/2}\,(Q=s,c,b)$,
and $x_{q,Q}$ being the bag eigenfrequencies.
$B$ is the bag constant, $n_q (n_{\qbar})$ and $n_Q (n_{\Qbar})$ 
are the lowest mode quark (antiquark) 
numbers for the quark flavors $q$ and $Q$ 
in the hadron $h$, respectively, 
and the $z_h$ parametrize the sum of the
center-of-mass and gluon fluctuation effects and are assumed to be
independent of density. Concerning the sign of $m_q^*$ in nuclear 
medium, it reflects nothing but the strength 
of the attractive scalar potential 
as in Eqs.~(\ref{diracu}) and~(\ref{diracd}), 
and thus naive interpretation of the mass for a (physical) particle, 
which is positive, should not be applied. 
The parameters are determined to
reproduce the corresponding masses in free space.
We chose the values,
$(m_q, m_s, m_c, m_b) = (5, 250, 1300, 4200)$ MeV for the
current quark masses, and $R_N = 0.8$
fm for the bag radius of the nucleon in free space. 
The quark-meson coupling constants, $g^q_\sigma$, $g^q_\omega$
and $g^q_\rho$, are adjusted to fit the nuclear 
saturation energy and density of symmetric nuclear matter, and the bulk
symmetry energy~\cite{Guichon,Guichonf,Saitof}. 
Exactly the same coupling constants, $g^q_\sigma$, $g^q_\omega$ and
$g^q_\rho$, are used for the light quarks in the mesons and baryons as in 
the nucleon. 

However, in studies of the kaon system, we found that it was
phenomenologically necessary to increase the strength of the vector
coupling to the non-strange quarks in the $K^+$ (by a factor of
$1.4^2$, i.e., $g_{K\omega}^q \equiv 1.4^2 g^q_\omega$) 
in order to reproduce the empirically extracted $K^+$-nucleus
interaction~\cite{Tsushimak}. This may be related to the fact that 
kaon is a pseudo-Goldstone boson, where treatment of the Goldstone 
bosons in a naive quark model is usually unsatisfactory. 
We assume this, $g^q_\omega \to 1.4^2 g^q_\omega$, 
also for the $D$, $\Dbar$~\cite{Tsushimad}, 
$B$ and $\Bbar$ mesons to allow an upper limit situation.
The scalar ($V^{h}_s$) and vector ($V^{h}_v$) potentials 
felt by the hadrons $h$,  
in nuclear matter are given by,
\bg
V^h_s &=& m^*_h - m_h,\,\,\label{spot}
\\
V^h_v &=&
  (n_q - n_{\bar{q}}) {V}^q_\omega + I^h_3 V^q_\rho,  
\qquad (V^q_\omega \to 1.4^2 {V}^q_\omega\,\,
{\rm for}\, K,\Kbar,D,\Dbar,B,\Bbar), 
\label{vpot}
\en
where $I^h_3$ is the third component of isospin projection  
of the hadron $h$. Thus, the vector potential felt by a heavy baryon 
with a charm or bottom quark, is equal to that of the hyperon with 
the same light quark configuration in QMC.

\begin{figure}[hbt]
\begin{center}
\epsfig{file=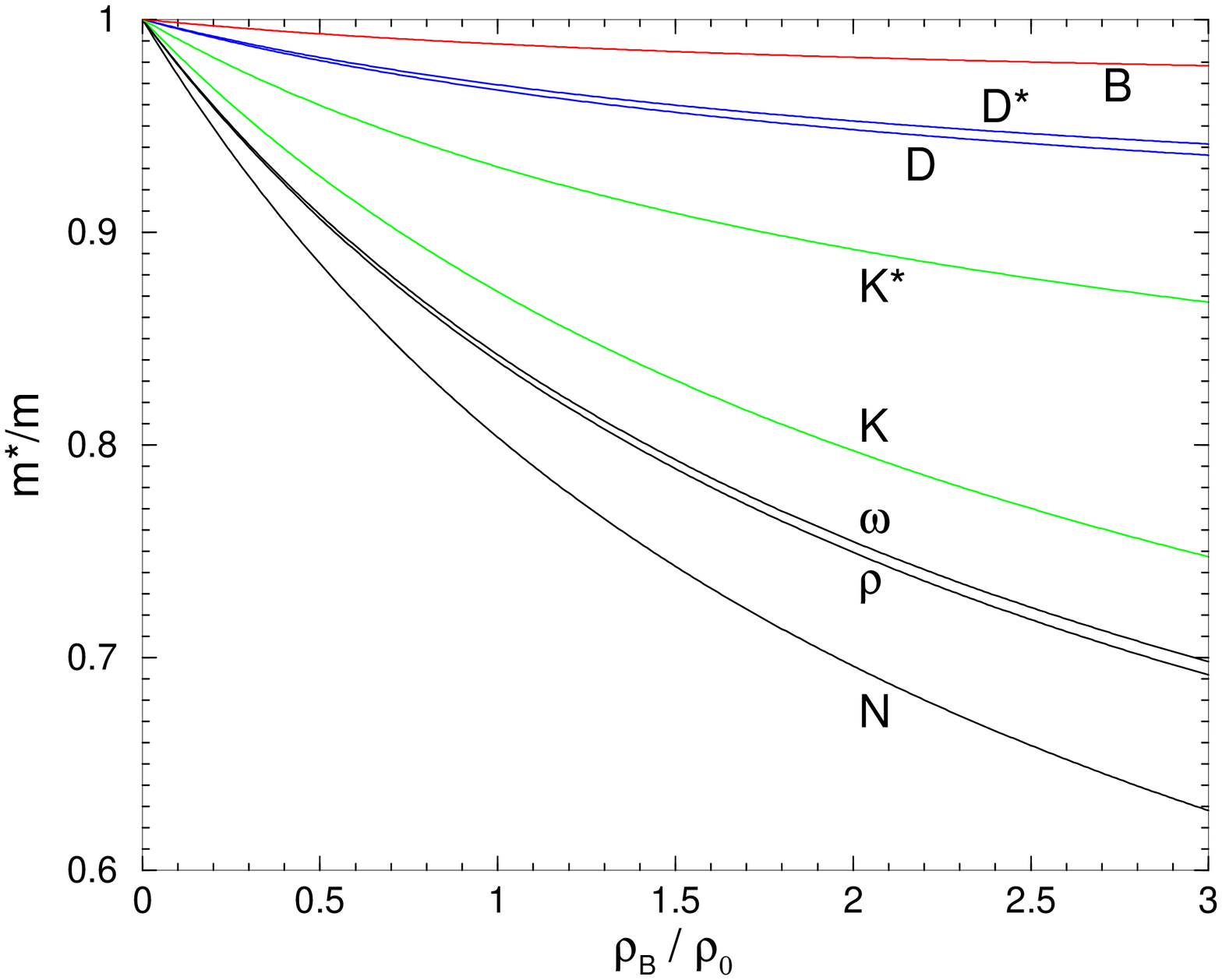,height=14cm,angle=-90}
\caption{Effective mass ratios for mesons in nuclear matter, where, 
$\rho_0 = 0.15$ fm$^{-3}$.
$\omega$ and $\rho$ stand for physical mesons which are treated in 
the quark model, and should not be confused with the fields appearing 
in the QMC model.
\label{mesonmass}
}
\end{center}
\end{figure}

\begin{figure}[hbt]
\begin{center}
\epsfig{file=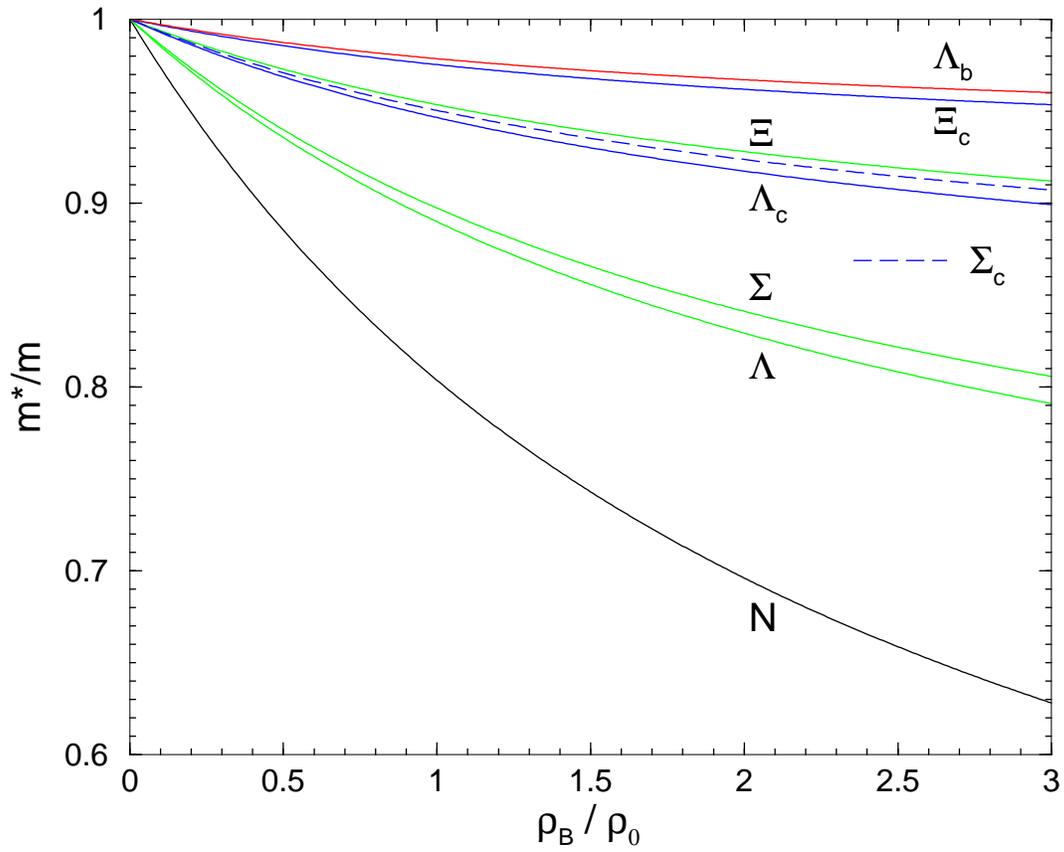,height=14cm,angle=-90}
\caption{Effective mass ratios for baryons in nuclear matter, where, 
$\rho_0 = 0.15$ fm$^{-3}$.
\label{baryonmass}
}
\end{center}
\end{figure}

In Figs.~\ref{mesonmass} and~\ref{baryonmass} we show ratios of 
effective masses (free masses + scalar potentials) 
versus those of the free particles, for 
mesons and baryons, respectively.
With increasing density the ratios decrease as usually expected, 
but decrease in magnitude is from larger to smaller: 
hadrons with only light 
quarks, with one strange quark, with one charm quark, and with one 
bottom quark. This is because their masses in free space 
are in the order from light to heavy. Thus, the net ratios for the 
decrease in masses (developing of scalar masses) compared to that of
the free masses becomes smaller. 
This may be regarded as a measure of the role of light 
quarks in each hadron system in nuclear matter, in a sense that by how 
much ratio do they lead to a partial restoration of the chiral 
symmetry in the hadron. In Fig.~\ref{mesonmass}, one can notice
somewhat anomalous behavior of the ratio for the kaon ($K$).
This is related to what we meant by the pseudo-Goldstone boson nature, 
i.e., its mass in free space is relatively light, $m_K \simeq 495$ MeV, 
and the ratio for the reduction in mass in nuclear medium is large.

\begin{figure}[hbt]
\begin{center}
\epsfig{file=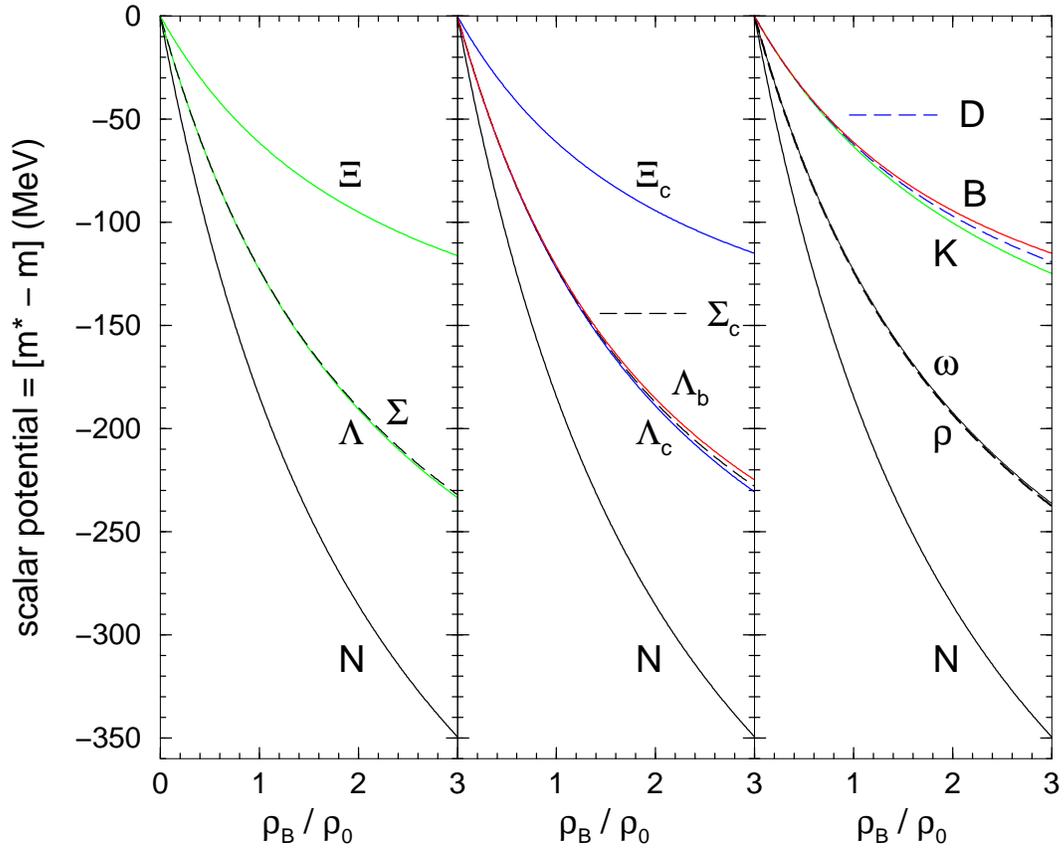,height=14cm,angle=-90}
\caption{Scalar potentials for various hadrons in nuclear matter, where, 
$\rho_0 = 0.15$ fm$^{-3}$.
(See also caption of Fig.~\ref{mesonmass}.)
\label{spotential}
}
\end{center}
\end{figure}

Perhaps it is much more quantitative and direct to compare  
scalar potentials of each hadron in the nuclear matter.
Calculated results are shown in Fig.~\ref{spotential}.
From the results it is confirmed that 
the scalar potential felt by the hadron $h$, $V_s^h$, 
follows a simple light quark number scaling rule:
\bge
V_s^h \simeq \frac{n_q + n_{\qbar}}{3} V_s^N, 
\ene
where $n_q$ ($n_{\qbar}$) is the number of light quarks (antiquarks) in 
the hadron $h$, and $V_s^N$ is the scalar potential felt by the nucleon. 
(See Eq.(\ref{spot}).)
It is interesting to notice that, the baryons with a charm and a bottom quark
($\Xi_c$ is a quark configuration, $qsc$), shows very similar 
features to those of hyperons with one or two strange quarks.
Then, we can expect that these heavy baryons with a charm or a bottom
quark, will also form charmed (bottom) hypernuclei, as the 
hyperons with strangeness do. (Recall that the repulsive, vector 
potentials are the same for the corresponding hyperons with the same light
quark configurations.)
Thus, an experimental investigation of such hypernuclei would be a fruitful
venture at JHF.

In addition, $B^-$ meson will also certainly form meson-nuclear bound states, 
because $B^-$ meson is $\ubar b$ and feels a strong attractive 
vector potential in addition to the attractive 
Coulomb force. This makes it much easier to be bound in a 
nucleus compared to the $D^0$~\cite{Tsushimad}, which is $c \ubar$ 
and is blind to the Coulomb 
force. 
This reminds us of a situation  
of the kaonic ($K^- (\ubar s)$) atom~\cite{katom1,katom2}.
A study of $B^- (\ubar b)$ atoms would be a fruitful experimental
program. Such atoms will have the 
meson much closer to the nucleus and will thus probe 
even smaller changes in the
nuclear density. This will be a complementary information to 
the $D^- (\bar{c} d)$-nuclear bound states, which would provide us an 
information on the vector potential in a nucleus~\cite{Tsushimad}.

To summarize, we have studied for the first time the properties 
of heavy baryons (hadrons) which contain a charm or a bottom quark 
in nuclear matter. Our results suggest that those heavy baryons will 
form charmed or bottom hypernuclei as was predicted in mid 70's.
We plan to report results for the
charmed and bottom hypernuclei studied quantitatively,
by solving a system equations for finite nuclei embedding a baryon
with a charm or a bottom quark~\cite{hypbc}.
In addition we can expect also $B^-$-nuclear bound (atomic) states 
based on the existing studies for the $D^0$ and kaonic atom.
Furthermore, formation of 
$B^-$-atoms would provide precise information 
on the nuclear density, which would be a complementary to 
that of the $D^-$-nuclear bound states.

\vspace{1cm}
\noindent
Acknowledgment\\
The authors would like to thank Prof. A.W. Thomas for the hospitality 
at the CSSM, Adelaide, where this work was initiated.
K.T. acknowledges support and warm hospitality at University of 
Alberta, where this work was completed.
K.T. is supported by the Forschungszentrum-J\"{u}lich, 
contract No. 41445282 (COSY-058). The work of F.K. is 
supported by NSERCC. 


%
\newpage
\newpage
\newpage

%
\end{document}